\documentstyle[aps,epsf]{revtex}
\begin{document}
\draft
\title{Fluctuation Effects in Steric Reaction-Diffusion Systems}
\author{Zoran Konkoli and Henrik Johannesson}
\address{
  {\em Institute of Theoretical Physics} \\
  {\em Chalmers University of Technology and G\"oteborg University}\\
  {\em S-412 96 G\"oteborg, Sweden}
}
\author{Benjamin P. Lee\footnote{Present address:
        Department of Physics, Bucknell University, Lewisburg, PA 17837}}
\address{
  {\em Polymers Division, National Institute of Science and Technology}\\
  {\em Gaithersburg, MD 20899}
}
\date{\today}
\maketitle
\begin{abstract}
We propose a simple model for reaction-diffusion systems 
with orientational constraints on the reactivity of particles, 
and map it onto a field theory with upper critical dimension 
$d_c=2$. To two-loop level the long-time particle density 
$N(t)$ is given by the {\em same} universal expression as for 
a nonsteric system, with $N(t) \sim t^{-d/2}$ for $d \le 2$. For 
slow rotations of the particles we find 
an intermediate regime where $N(t) \sim t^{-d/4}$, 
with the crossover to the nonsteric asymptotics determined by the 
rates of rotations and reactions. Consequences for 
experiments are discussed.
\end{abstract}

Diffusion-controlled reactions are ubiquitous in Nature and span a wide 
range of phenomena --- from chemical catalysis in cell biology to 
matter-antimatter annihilation in the early Universe. In these processes 
the time $t_D$ for a particle to diffuse 
to its nearest neighbor is much larger than the characteristic time $t_R$
for the particles to react, once in proximity. As a result, large density fluctuations
may develop at late times, resulting in   
anomalous kinetics with universal scaling behavior 
\cite{mikhailov}. 

Our understanding of diffusion-controlled reactions has advanced 
significantly in the last few years. Results obtained by new theoretical approaches (like
renormalization group methods  
\cite{leecardy} or mappings to integrable models
\cite{alcaraz}) can now be tested against high-quality experimental data, 
including measurements on low-dimensional processes where effects from fluctuations 
are most pronounced \cite{kroon}. However, 
little attention has been paid to the modeling of fluctuation effects in {\em steric} 
reaction-diffusion systems, i.e. systems where the reactivity between 
particles depend on their relative orientation. Examples include 
diffusion-controlled enzyme reactions 
\cite{schoup}, effective reaction-diffusion models of protein folding 
\cite{karplus}, as well as generic solution kinetics of 
molecules with nonuniform surface reactivity \cite{solc}.  
Here the chemically active 
part of a reactant is located in a specific region, usually represented 
by an {\em active spot} \cite{solc} on the sphere modeling the reactant. 
Two particles react only if their active spots touch, signaling a 
successful ``docking'' for reaction. Despite the importance of this class of processes, 
the theoretical effort so far has been limited to estimating 
how the reaction rates effectively decrease due to orientational constraints 
\cite{solc,baldo}. The lack of a more comprehensive theory 
reflects the subtlety of the 
problem, which here makes a faithful treatment quite difficult. 

In this Letter we take a new route and propose a 
{\em minimal} model for steric reaction-diffusion systems, specifically 
annihilation processes.
This model - in spirit an ``Ising-type'' model for steric reaction-diffusion - 
has the advantage that it can be analyzed by field-theoretic renormalization group 
(RG) techniques, allowing for precise predictions about the long-time kinetics.  
In spite of its simplicity the model contains some very interesting 
information. In particular, for sufficiently slow internal rotations of the particles  
our results reveal that steric reactivity in dimensions $d \le 2$ drives a
crossover from a density decay $N(t) \sim t^{-d/4}$ at large intermediate times  
to $N(t) \sim t^{-d/2}$ at later times (with logarithmic corrections in $d=2$). 

To define the model we partition configuration space into cells
$\Delta {\bf r}$, and time into intervals $\Delta t$. The particles are
represented by spheres with an active spot that is either ``up'' or
``down'' with respect to a fixed reference direction, and ``A'' (``B'') denotes a
sphere with the spot up (down) \cite{lowdim}. Thus, the orientation
of a sphere is restricted to two states, A and B. The number of $A$
and $B$ particles in the cells at a time step $n\Delta t, n=1,2,....$, are given
by the sets of cell occupation numbers $\{a\}_{n\Delta t}$ and
$\{b\}_{n\Delta t}$ respectively. 
We assume that the microscopic dynamics lead to a {\em Markov process} \cite{vankampen}
with waiting time interval $\Delta t$, and assign probabilities that within $\Delta t$ a particle
in a given cell will {\em react} (if the cell is nonempty), {\em jump} to a neighboring cell {\em
with} or {\em without rotation} (with rates $\tilde{D}_{transl + rot}$ and $\tilde{D}$
respectively), or {\em stay} in the cell {\em with rotation} (with rate $\tilde{D}_{rot}$) or {\em
without}. 
In the continuum limit $\tilde{D}_{rot}$
and $\tilde{D}_{trans+rot}$ fold into one effective parameter 
which we refer to as $D_{rot}$, with the continuum limit translational diffusion constant
denoted by $D$.
At first disregarding the influence of microscopic rotations we
assume that two particles in a cell may react and annihilate with some
rate $\delta_0$ only if one is A and the other B, with $\delta_0$ an
average over all possible active (``spot-to-spot'') and nonactive
encounters within the time interval $\Delta t$.
Since microscopic rotations in the time interval $\Delta t$ may  turn a nonactive
encounter into an active one we also assign a nonzero reaction rate $\lambda_0$ to the cases
where both particles initially are of A {\em or} B type.
The assignment of values to the rate constants is immaterial, but
clearly $\lambda_0$ will approach $\delta_0$ for  sufficiently
fast microscopic rotations. 
Thus, to ``optimize'' effects from steric reactivity we assume that
$\lambda_0 \ll \delta_0$, and also $D_{rot}/D \ll
1$. This limit of ``slow'' rotations
is indeed realized in many experimental systems \cite{baldo}.

The model thus constructed is equivalent to a generalized two-species 
annihilation system where $A$ and $B$ particles diffuse by  
translations  and ``flips'' $A \leftrightarrow B$  
and react according to \cite{howard}:
\begin{equation}
  A+A \stackrel{\lambda_0}{\longrightarrow}\emptyset   \ \ \ \ \
  A+B \stackrel{\delta_0 }{\longrightarrow}\emptyset   \ \  \ \ \
  B+B \stackrel{\lambda_0}{\longrightarrow}\emptyset . 
  \label{reaction}
\end{equation}
Without rotations $(\lambda_0 = 0$ and $D_{rot} =0)$ one 
immediately 
obtains the well-known $A+B \rightarrow 0$ universality class for the 
asymptotic mean density: $n_a(t)=n_b(t) \sim t^{-d/4}$ 
\cite{TouissantWilczek,leecardy}. To explore the more realistic case including  
rotations we consider the master equation 
\begin{equation}
  \frac{d}{dt} P(c,t) = \sum_{c'}R_{c' 
  \rightarrow c} P(c',t)  
- \sum_{c'}R_{c 
  \rightarrow c'} P(c,t). \label{aaMEQ}
\end{equation}
Here $P(c,t)$ denotes the 
probability of a given configuration $c \equiv (\{a\}, \{b\})$ at time $t$, and $R_{c' 
\rightarrow c}$ is the transition rate from state 
$c' \equiv (\{a\}',\{b\}')$ into $c$, determined by $D, D_{rot}, \delta_0$ and 
$\lambda_0$. We take $P(c,0)$ as a Poisson distribution, with averages  
denoted by $n_{a,0}$ and $n_{b,0}$ for $A$ and $B$ particles respectively. In the absence of 
anisotropies or external fields we expect that $n_{a,0} = n_{b,0}$, but for now  
leave the initial densities unspecified. 

Following Refs. \cite{Doi,Pel1}, the master equation (2) 
can be mapped 
onto a field theory in the continuum limit, with A and B particles described by 
scalar fields $a$ and $b$ respectively. Introducing  
$\phi \equiv (a+b)/2 \:, \psi \equiv (a-b)/2 \:$ (with conjugated fields defined without the
factor of $1/2$), and corresponding
initial densities $n_{\phi,0} \equiv (n_{a,0}+n_{b,0})/2 \:, n_{\psi,0} \equiv 
(n_{a,0}-n_{b,0})/2$, we obtain the action
\begin{eqnarray}
  S &=& \int d^dx \int dt [ 
       \bar{\phi}(\partial_t-\nabla^2)\phi
       + \bar{\psi}(\partial_t-\nabla^2)\psi + \rho\bar{\psi}\psi
       +(\lambda_0+\delta_0)\bar{\phi}\phi^2 
       +(\lambda_0-\delta_0)\bar{\phi}\psi^2             \cr
   & &  +2\lambda_0\bar{\psi}\psi\phi               
     +\frac{\lambda_0+\delta_0}{4}\bar{\phi}^2\phi^2
      +\frac{\lambda_0-\delta_0}{4}\bar{\phi}^2\psi^2
      +\frac{\lambda_0-\delta_0}{4}\bar{\psi}^2\phi^2
      +\frac{\lambda_0+\delta_0}{4}\bar{\psi}^2\psi^2    \cr
   & & +\lambda_0\bar\phi\bar\psi\phi\psi ] 
        -\int d^dx ( n_{\phi,0} \bar{\phi}(x,0) 
      + n_{\psi,0} \bar{\psi}(x,0) ) ,           \label{Spp}
\end{eqnarray}
where we have rescaled the time variable, $Dt \rightarrow t$, and where 
$\rho  \sim D_{rot}/D$. Using 
(\ref{Spp}), the mean densities can be calculated as $n_c(t) = \langle c(t)\rangle_S \ 
(c=a,b)$, where $\langle \ \rangle
_S$ denotes an average with respect to the action $S$.
The rate constants, here playing the role of coupling 
constants, become dimensionless when $d=2$, suggesting this as the upper 
critical dimension above which mean field theory (``normal reaction 
kinetics'' \cite{mikhailov}) should be valid, a result confirmed by standard power counting
techniques \cite{Amit}. Hence, we concentrate on $d 
\le 2$. These dimensions can be explored by using 
RG improved perturbation theory in  
$d_c = 2$ and then continuing the results to $d < 2$ via an 
$\epsilon$-expansion. As seen in Eq. (\ref{Spp}), rotational diffusion acts as a mass 
perturbation in the $\psi$-sector, and for simplicity we here focus on the 
``strong'' steric limit where this term is omitted, commenting on the general
case as we go along.

Using dimensional regularization, divergences in perturbation theory show 
up as poles in $\epsilon = 2-d$. Analogous to the single-species case 
$A+A \rightarrow 0$ \cite{LeeAA}, these poles can be removed by simply 
renormalizing the rate constants: $\lambda_0 \rightarrow \lambda_R$ and 
$\delta_0 \rightarrow \delta_R$. The renormalized rate constants are defined by primitively divergent
vertex functions evaluated at an arbitrary momentum scale $\kappa$, which is further used to
define
dimensionless bare and renormalized rates $g_0 \equiv \kappa^{-\epsilon}\delta_0, 
h_0 \equiv \kappa^{-\epsilon} 
\lambda_0$ and $g_R \equiv \kappa^{-\epsilon} \delta_R, h_R \equiv \kappa^{-\epsilon} \lambda_R$, 
respectively.  
These are related by the exact 
result
\begin{equation}
 g_R = \frac{g_0}{1+g_{0}/g^*} , \ \ \ \ \ h_R = \frac{h_0}{1+h_0/h^*}, \ \ \ \ \
                                     \label{ren}
\end{equation}
with $g^* = h^* = \Gamma(\epsilon/2)^{-1}(8\pi)^{d/2}$ the fixed point 
to which both rate constants flow.  
We are interested in the 
large-time mean densities $n_a(t)$ and $n_b(t)$ which depend in principle on the 
initial densities $n_{a,0}$ and $n_{b,0}$, and the bare rates $\lambda_0$ and 
$\delta_0$, expressed through $h_R, g_R$ and $\kappa$. 
However, the densities cannot depend on the arbitrarily chosen  
scale $\kappa$, and this implies a Callan-Symanzik equation
\begin{equation}
 n_c(t;g_R,h_R;n_{a,0},n_{b,0};\kappa)=
     (\kappa^2t)^{-d/2} n_c(\kappa^{-2};
       \tilde{g}_R(t),\tilde{h}_R(t);
        \tilde{n}_{a,0}(t),\tilde{n}_{b,0}(t);\kappa)     \:,\: \
  c=a,b                                           \label{CSAB}
\end{equation}
with running coupling constants 
\begin{equation} 
  \tilde{g}_R(t) = 
   \frac{g^*}{1+(t/t_\delta^*)^{-\epsilon/2}}, \ \ \ \tilde{h}_R(t) =
   \frac{h^*}{1+(t/t_\lambda^*)^{-\epsilon/2}}, 
   \label{run}
\end{equation}
and densities $\tilde{n}_{c,0}(t)=(\kappa^2t)^{d/2}n_{c,0}$, with $c=a,b$.
The crossover times are given by $t_\delta^* = (g^*/\delta_0)^{2/\epsilon}$ and 
$t_\lambda^* = (h^*/\lambda_0)^{2/\epsilon}$  
respectively.
The fact that the rate constants flow to the {\em same} fixed point 
coupling $g^* = h^*$ has some unexpected 
consequences, and dramatically simplifies calculations as compared to the 
two-species case $A+B \rightarrow 0$ with 
$\lambda_0 = 0$ --- a notoriously difficult problem within the RG \cite{leecardy}. 
  
To calculate the mean densities $n_a(t)$ and $n_b(t)$, the perturbation series has to 
be organized in such a way as to make it well-behaved when the initial 
running densities flow to infinity in Eq. (\ref{CSAB}). At tree level this 
can be done by simply ``dressing'' the initial densities $n_{c,0}$ ($c=a,b$)  
\begin{equation}
  \epsfxsize=6cm
  \raisebox{-0.5cm}{
    \epsfbox[59 339 553 452]{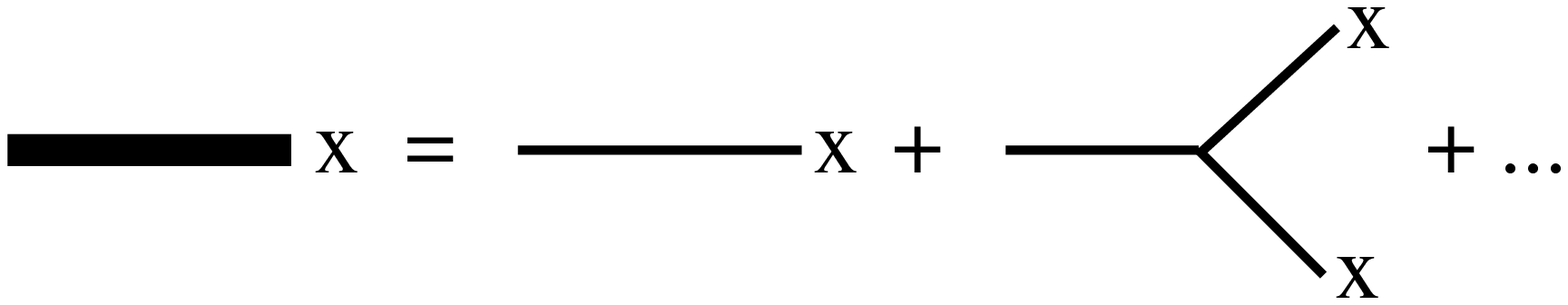}
  }
  \label{ncl}
\end{equation}
{\noindent}and free propagators $G_0(t_1-t_2;p) = 
\theta(t_1-t_2)e^{-p^2(t_1-t_2)}$
\begin{equation}
  \epsfxsize=6cm
  \epsfbox[85 370 526 436]{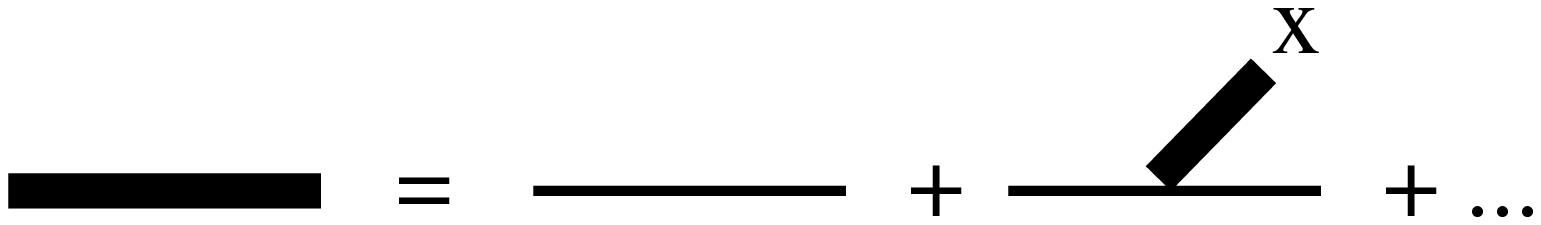}
  \label{Gcl}
\end{equation}
{\noindent }In these diagrams a thin (thick) line corresponds to a free (dressed) 
propagator, with a cross representing an initial density. The dressed 
tree-level densities $\langle a(t)\rangle$ and 
$\langle b(t)\rangle$ (with averages taken only over tree-level diagrams) 
are solutions of the coupled {\em classical rate 
equations}  
\begin{equation}
\frac{d}{dt}\langle a(t)\rangle = -(\lambda_0 \langle a(t)\rangle^2 + 
\delta_0 \langle a(t)\rangle \langle b(t)\rangle) , \ \ \ \ \ 
  \frac{d}{dt}\langle b(t)\rangle = -(\lambda_0 \langle b(t)\rangle^2 + 
\delta_0 \langle b(t)\rangle \langle a(t)\rangle) ,      \label{mf}
\end{equation}
which for equal initial densities $n_{a,0}=n_{b,0}=n_0$ yield the 
prediction
\begin{equation}
\langle a(t)\rangle=\langle b(t)\rangle=\frac{n_0}{1+n_0(\lambda_0+\delta_0)t}.  \label{smfphi}
\end{equation}

This result is drastically changed by fluctuations, as revealed by a loop 
expansion of the densities. To carry it out we need the dressed tree-level 
propagators, which for equal initial densities can be calculated exactly. 
In the $\phi$-$\psi$ basis:
\begin{eqnarray}
  G_\xi(p,t_1-t_2)&=&
        \theta(t_1-t_2)e^{-p^2(t_1-t_2)}
   \left( \frac{1+n_0(\lambda_0+\delta_0)t_2}
               {1+n_0(\lambda_0+\delta_0)t_1}
   \right)^{\Delta_{\xi}}, \ \ \ \xi=\phi, \psi              \label{Gxi} 
\end{eqnarray}
with $\Delta_\phi=2$ and $\Delta_\psi=2\lambda_0/(\lambda_0 + \delta_0)$.
At one-loop level there are two diagrams which contribute to $\langle\phi\rangle$ (with
$\langle \ \rangle$ now denoting a full average):
\begin{equation}
  \langle\phi\rangle_{1-loop}
  \epsfxsize=6cm
  \epsfbox[72 382 540 445]{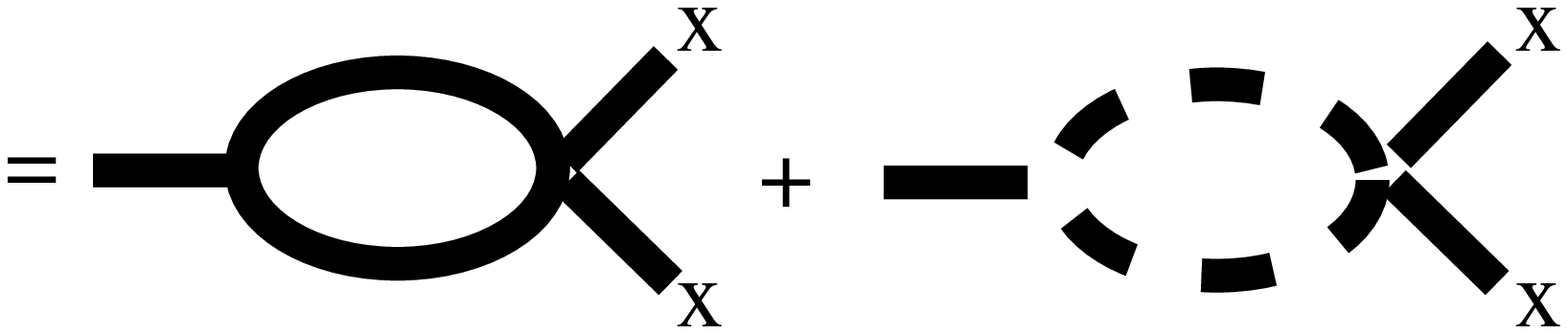}
\end{equation}

{\noindent}with a full (dashed) line denoting a ${\phi} ({\psi})$ propagator.
After performing the momentum loop integrations and 
taking $n_0 \rightarrow \infty$, one obtains 
\begin{equation}
  \langle\phi\rangle_{1-loop}= \frac{1}{2} (8\pi t)^{-d/2}
    \int_{0}^{1}dx x^{1-d/2}\int_{0}^{1}dy[y^2+z^2y^{2z}]
      (1-y)^{-d/2}
                                              \label{phi1Lz}
\end{equation}
with $t_1\equiv tx$ and $t_2 \equiv txy$, and with $z \equiv (\lambda_R 
- \delta_R)/(\lambda_R + \delta_R)$. 
Eqs. (\ref{ren}) and (\ref{run}) 
imply that $z \rightarrow 0$ for large times, thus killing the second 
term in the square bracket in Eq. (\ref{phi1Lz}). By an 
$\epsilon$-expansion, and using that $\langle \psi \rangle =0$ for equal initial densities, 
we thus obtain for large times: 
\begin{equation}
  n_a(t) = n_b(t) = \frac{1}{2} \left[
    \frac{1}{4\pi\epsilon}+\frac{\ln 8\pi-\frac{5}{2}}{8\pi}
     +\ldots \right]t^{-d/2} , \ \ \ \ t \rightarrow \infty, \ \ \ \lambda \neq 0,
\ \ \ d<2   \label{A1}
\end{equation} 
This result, which we have checked to two-loop level,  yields the {\em same} universal 
expression for the total density $N(t) = n_a(t) + n_b(t)$ as for a 
nonsteric system \cite{LeeAA}. The inclusion of rotational diffusion with 
$\rho \neq 0$ in Eq. (\ref{Spp}) does not influence the expression 
in Eq. (\ref{A1}): since $\rho$ does not renormalize, the long-time
asymptotics of the theory stays the same \cite{kjl}.
The fact that the asymptotic density is blind to steric 
effects and exhibits not only the same decay exponent $-d/2$ as a 
nonsteric system but also 
the same {\em amplitude} is a striking manifestation of universality in 
reaction-diffusion processes. 
 
Eq. (\ref{phi1Lz}) reveals how $\lambda_0$ 
controls the universality class of the model. In addition to the 
pole in $1/\epsilon$, there 
is a second {\em potential} divergence coming from the $y$-integration 
due to the flow $n_0 \rightarrow \infty$ under the RG. 
However, as $n_0$ flows,
$z$ flows as well {\em unless} $\lambda_0=0$,
and the asymptotic value of $z$  
determines the fate of the integral: for $z>-1/2$ ($z \le - 1/2$) the 
$y-$integral is finite (divergent).       
If $\lambda_0=0$ one has the $A+B \rightarrow 0$ universality class 
\cite{leecardy}. For this case $z=-1<-1/2$ for all times and 
the integral diverges.  
On the other hand, for {\em any} $\lambda_0\ne 0$, $z 
\rightarrow 0$ and accordingly $z>-1/2$ for large times, producing a 
well-behaved expansion.   
This gives the $A+A \rightarrow 0$ universality class of Eq. (\ref{A1}). 
Our results imply that even if $\lambda_0$ is arbitrarily small, the 
system will eventually behave as if $\lambda_0=\delta_0$ {\em (non-steric 
limit)}. How long does it take for this to happen? 
Since $z=-1/2$ is the 
critical value which controls the 
$n_0$ dependence of perturbation theory, and thus the crossover to the 
$A+A \rightarrow 0$ universality class, it is natural to define a 
crossover time $t_z$ such that $z(t_z)=-1/2$. Eq. (\ref{run})
together 
with the definition of $z$ shows that
\begin{equation}
  t_z=\left[ \frac{g^*}{2} 
        \left( 
          \frac{1}{\lambda_0} - \frac{3}{\delta_0} 
        \right)
      \right]^{2/\epsilon}                        \label{tz}
\end{equation}
Since $t^*_{\delta} \ll t_z \sim t^*_{\lambda}$ for small $\lambda_0$, 
there is a large intermediate time interval with a strong $n_0$- 
dependence where $\tilde{g}_R$ is close 
to its fixed point value $g^*$, while $\tilde{h}_R$ is still
small. A strong $n_0$-dependence with $\tilde{g}_R \sim g^*$ implies scaling 
with the exponent $-d/4$ when $\tilde{h}_R =0$ (no 
rotations) \cite{leecardy}. Since a small $\tilde{h}_R$ cannot change this 
behavior, it will be present also for slow rotational modes with $\lambda_0 \neq 0$. 
The inclusion of rotational diffusion with $\rho \neq 0$ in Eq. 
(\ref{Spp}) will decrease $t_z$, but for sufficiently small $\rho$ there 
will still be an intermediate regime with a strong $n_0$-dependence, 
supporting scaling with the exponent $-d/4$ \cite{kjl}.   
We expect this property to hold generically, also for more realistic 
systems with continuous degrees of freedom. By analogy, we also expect a 
similar crossover from $n(t) \sim t^{-d/4}$ to the usual mean-field behavior  
$n(t) \sim t^{-1}$ for steric systems in dimensions $d > 2$. Work on this problem 
is in progress.

What causes the crossover from the $t^{-d/4}$ decay to the $t^{-d/2}$ 
deacy for small $\lambda_0$? The most probable scenario is that $A$ and $B$ 
particles begin to segregate into A and B rich regions, a process during 
which the $A+B\rightarrow 0$ reaction is dominant ($-d/4$ decay exponent). 
Once A and B regions are formed, however, the $A+B\rightarrow 0$ reaction 
becomes unimportant and the $A+A\rightarrow 0$ and $B+B\rightarrow 0$ reactions 
become dominant despite the small value of 
$\lambda_0$, resulting in the $-d/2$ decay exponent. 

Figure 1 shows the results of a numerical simulation
of the master equation
(\ref{aaMEQ}) in $d=1$ \cite{numerics}. The uppermost curve corresponds to
$\lambda=0$ and $\delta=10$ ($A+B\rightarrow 0$ model with decay exponent
$-d/4$), while the lowest curve corresponds to the case with equal
reaction rates $\lambda=\delta=10$, representing the asymptotic limit of
equal rate constants as given in (\ref{run}). As predicted, for small
values of $\lambda$ (intermediate curves) the decay exponent (slope of the
curves)  changes from $-d/4$ for early times to $-d/2$ at late times, with
crossovers consistent with (\ref{tz}) (scaling with $1/\lambda^2$ for
small $\lambda$). 

The present model can be generalized to an arbitrary number of
orientations $N$ by introducing the ''particles'' $A_1,...,A_N$, and
reactions $A_\alpha+A_\beta
\stackrel{\lambda_{\alpha,\beta}}{\longrightarrow}\emptyset$. This
generalized system can be shown to belong to the same universality class
as the present $N=2$ model as long as all rates $\lambda_{\alpha,\beta}$
are nonzero \cite{kjl}. However, if one or several of the rates
$\lambda_{\alpha,\beta}$ vanish, a different scaling behavior results
(most probably the $t^{-d/4}$ asymptotics of the "A+B" model, although
this is still to be established). 

The recent breakthrough in preparing and monitoring low-dimensional
chemical reactions --- using carbon nanotubes as effectively
one-dimensional ``test tubes'' \cite{science,nature} --- opens up the
possibility to experimentally test our predictions. The essential elements
to be present in a laboratory set-up are that (i) the reactions are
sufficiently fast to ensure the diffusion-controlled condition $t_R <<
t_D$, (ii) the reactants have a highly anisotropic surface reactivity with
sufficient slow rotational diffusion to make the system strongly steric,
and (iii) the reactive products precipitate out of the system on a short
time scale compared to $t_D$ and hence do not influence the kinetics
\cite{brownbook}. An observation of the crossover behavior predicted in
the present work would be extremely interesting as it would exhibit, for
the first time, a reaction-diffusion process supporting two distinct time
scales with different anomalous scaling. 

We thank M. Howard, Z. R\'{a}cs, G. Sch\"{u}tz and particularly U. T\"{a}uber for 
enlightening discussions, and R. 
Sprik and D. Ugarte for correspondence about the experimental background. Support from 
the 1997 Budapest Workshop on Nonequilibrium Phenomena (Z.K.),
the Swedish Natural Science Research Council (H.J.), and a National
Research Council Research Associateship (B.P.L.) are acknowledged.


\begin{figure}
\caption{Density deacy obtained from a numerical simulation of the master equation in (2) in $d=1$ with
$10^5$ cells. Here $\delta=10$ and $n_A^0=n_B^0=3.5$ particles/site. 
}
\end{figure}


\begin{thebibliography}{19}
\bibitem{lowdim} In low dimensions consistency within the model requires the reference axis 
to be chosen parallel with the line (plane) defining the d=1 (d=2) manifold.
\end{thebibliography}
\end{document}